\documentclass[aps,prc,twocolumn,showpacs,amsmath,amssymb,superscriptaddress]{revtex4-1}

\usepackage{graphicx}
\usepackage{dcolumn}
\usepackage{bm}
\usepackage{color}

\begin{document}

\title{Performance of the upgraded ultracold neutron source at Los
  Alamos National Laboratory and its implication
  for a possible neutron electric dipole moment experiment}

\author{T.~M.~Ito}
\email[Corresponding author. Electronic address: ]{ito@lanl.gov}
\affiliation{Los Alamos National Laboratory, Los Alamos, New Mexico 87545, USA}

\author{E.~R.~Adamek}
\affiliation{Indiana University, Bloomington, Indiana, 47405, USA}

\author{N.~B.~Callahan}
\affiliation{Indiana University, Bloomington, Indiana, 47405, USA}

\author{J.~H.~Choi}
\affiliation{North Carolina State University, Raleigh, North Carolina,
  27695, USA}

\author{S.~M.~Clayton}
\affiliation{Los Alamos National Laboratory, Los Alamos, New Mexico 87545, USA}

\author{C.~Cude-Woods}
\affiliation{Los Alamos National Laboratory, Los Alamos, New Mexico 87545, USA}
\affiliation{North Carolina State University, Raleigh, North Carolina,
  27695, USA}

\author{S.~Currie}
\affiliation{Los Alamos National Laboratory, Los Alamos, New Mexico 87545, USA}

\author{X.~Ding}
\affiliation{Virginia Polytechnic Institute and State University,
  Blacksburg, Virginia, 24061, USA}

\author{D.~E.~Fellers}
\affiliation{Los Alamos National Laboratory, Los Alamos, New Mexico 87545, USA}

\author{P.~Geltenbort}
\affiliation{Institut Laue-Langevin, 38042, Grenoble Cedex 9, France}

\author{S.~K.~Lamoreaux}
\affiliation{Yale University, New Haven, Connecticut, 06520, USA}

\author{C.~Y.~Liu}
\affiliation{Los Alamos National Laboratory, Los Alamos, New Mexico 87545, USA}
\affiliation{Indiana University, Bloomington, Indiana, 47405, USA}

\author{S.~MacDonald}
\affiliation{Los Alamos National Laboratory, Los Alamos, New Mexico 87545, USA}

\author{M.~Makela}
\affiliation{Los Alamos National Laboratory, Los Alamos, New Mexico 87545, USA}

\author{C.~L.~Morris}
\affiliation{Los Alamos National Laboratory, Los Alamos, New Mexico 87545, USA}

\author{R.~W.~Pattie~Jr.}
\affiliation{Los Alamos National Laboratory, Los Alamos, New Mexico 87545, USA}

\author{J.~C.~Ramsey}
\affiliation{Los Alamos National Laboratory, Los Alamos, New Mexico 87545, USA}

\author{D.~J.~Salvat}
\affiliation{Indiana University, Bloomington, Indiana, 47405, USA}

\author{A.~Saunders}
\affiliation{Los Alamos National Laboratory, Los Alamos, New Mexico 87545, USA}

\author{E.~I.~Sharapov}
\affiliation{Joint Institute of Nuclear Research, Dubna, Moscow
  Region, Russia, 141980}

\author{S.~Sjue}
\affiliation{Los Alamos National Laboratory, Los Alamos, New Mexico 87545, USA}

\author{A.~P.~Sprow}
\affiliation{University of Kentucky, Lexington, KY 40506, USA}

\author{Z.~Tang}
\affiliation{Los Alamos National Laboratory, Los Alamos, New Mexico 87545, USA}

\author{H.~L.~Weaver}
\affiliation{Los Alamos National Laboratory, Los Alamos, New Mexico 87545, USA}

\author{W.~Wei}
\affiliation{Los Alamos National Laboratory, Los Alamos, New Mexico 87545, USA}

\author{A.~R.~Young}
\affiliation{North Carolina State University, Raleigh, North Carolina,
  27695, USA}


\date{\today}

\begin{abstract}
The ultracold neutron (UCN) source at Los Alamos National Laboratory
(LANL), which uses solid deuterium as the UCN converter and is driven
by accelerator spallation neutrons, has been successfully operated for
over 10 years, providing UCN to various experiments, as the first
production UCN source based on the superthermal process. It has
recently undergone a major upgrade. This paper describes the design
and performance of the upgraded LANL UCN source. Measurements of the
cold neutron spectrum and UCN density are presented and compared to
Monte Carlo predictions. The source is shown to perform as modeled.
The UCN density measured at the exit of the biological shield was
$184(32)$~UCN/cm$^3$, a four-fold increase from the highest previously
reported. The polarized UCN density stored in an external chamber was
measured to be $39(7)$~UCN/cm$^3$, which is sufficient to perform an
experiment to search for the nonzero neutron electric dipole moment
with a one-standard-deviation sensitivity of $\sigma(d_n) = 3\times
10^{-27}$~$e\cdot$cm.
\end{abstract}


\maketitle

Ultracold neutrons (UCN)~\cite{Ign90,Gol91} are defined to be neutrons
of sufficiently low kinetic energies that they can be confined in
material and magnetic bottles, corresponding to kinetic energies below
about 340~neV. UCN are playing increasingly important roles in the
study of fundamental physical interactions (see
e.g. Refs.~\cite{Dub11,You14}). Searches for a non-zero electric
dipole moment of the neutron (nEDM), which are performed almost
exclusively using UCN~\cite{Khr97,Lam09,ILL,Ser15}, probe new sources of time
reversal symmetry violation~\cite{Pos05,Eng13} and may give clues to the
puzzle of the matter-antimatter asymmetry in the
Universe~\cite{Cir10,Mor13}. The free neutron lifetime, which is
measured using UCN~\cite{Ser08,Pat17} or beams of cold
neutrons~\cite{Yue13}, is an important input parameter needed to
describe Big-Bang nucleosynthesis. Measurements of neutron decay
correlation parameters performed using
UCN~\cite{Pat09,Liu10,Pla12,Men13,Bro17} as well as cold
neutrons~\cite{Mun13}, along with measurement of the free neutron
lifetime, test the consistency of the standard model of particle
physics and probe what may lie beyond it~\cite{Cir13}. Precision
studies of bound quantum states of the neutron in gravitational fields
are performed to search for new interactions~\cite{Jen11}.

For decades, the turbine UCN source~\cite{Ste86} at the PF2 Facility
of Institut Laue-Langevin (ILL) provided UCN to various experiments as
the world's only UCN source with sufficient UCN density and flux.
Ultimately the performance of the UCN experiments performed there was
limited by the available UCN density and flux.  This led to
development of many new UCN sources around the world based on the
superthermal process~\cite{Gol75} in either liquid helium
(LHe)~\cite{Gol77} or solid deuterium (SD$_2$, where `D' denotes
deuterium `$^2$H' )~\cite{Gol83,Yu85,Mor03,Sau04} coupled to
spallation or reactor neutrons. See e.g. Ref.~\cite{You14} for a list
of operational and planned UCN sources around the world.

At Los Alamos National Laboratory (LANL) a UCN source based on a
SD$_2$ converter driven by spallation neutrons has been operated
successfully for over 10 years~\cite{Sau13}. This was the first
production UCN source based on superthermal UCN production. As the
only operational UCN source in the US and as one of the two
multi-experiment UCN facilities in the world (along with the ILL
turbine source), it has provided UCN to various experiments including
the UCNA~\cite{Pat09,Liu10,Pla12,Men13,Bro17}, UCNB~\cite{Bro16}, and
UCN$\tau$~\cite{Pat17} experiments as well as development efforts for
the nEDM and Nab experiments at SNS~\cite{Tan16, Bro16}.

This source has recently undergone a major upgrade, primarily
motivated by the desire to perform a new nEDM experiment with improved
sensitivity~\cite{Ito14}. The current upper limit on the nEDM, set by
an experiment performed more than a decade ago at the ILL turbine UCN
source, is $d_n < 3.0\times 10^{-26}$~$e\cdot$cm (90\%
C.L.)\cite{ILL}, which was statistics limited. Further
improvement on the sensitivity of experiments to the nEDM, currently
attempted by many efforts worldwide, has been hindered by the lack of
sufficiently strong sources of UCN, although it has been shown through
various efforts that the necessary control for known systematic
effects at the necessary level is likely to be achievable. An
estimate~\cite{Ito14} indicated that an upgrade of the LANL UCN source
would provide a sufficient UCN density for an nEDM experiment with a
sensitivity of $\sigma(d_n) \sim 3 \times 10^{-27}$~$e\cdot$cm, which
formed the basis of the source upgrade reported here.

The basic design of the source, which was unchanged through the
upgrade, is as follows. Spallation neutrons produced by a pulsed
800-MeV proton beam striking a tungsten target were moderated by
beryllium and graphite moderators at ambient temperature and further
cooled by a cold moderator that consisted of cooled polyethylene
beads. The cold neutrons were converted to UCN by downscattering in an
SD$_2$ crystal at 5~K. UCN were directed upward 1~m along a vertical
guide coated with $^{58}$Ni, to compensate for the 100-neV boost that
UCN receive when leaving the SD$_2$, and then 6~m along a horizontal
guide made of stainless steel (and coated with nickel phosphorus for
the upgraded source) before exiting the biological shield. At the
bottom of the vertical UCN guide was a butterfly valve that remained
closed when there was no proton beam pulse striking the spallation
target, in order to keep the UCN from returning to the SD$_2$ where
they would be absorbed.

When in production, the peak proton current from the accelerator was
typically 12~mA, delivered in bursts of 10~pulses each 625-$\mu$s long
at 20~Hz, with a gap between bursts of 5 s. The total charge delivered
per burst was $\sim$45~$\mu$C in 0.45~s. The time averaged current
delivered to the target was $\sim$9~$\mu$A.

\begin{figure}
\centering
\includegraphics[width=3.5in]{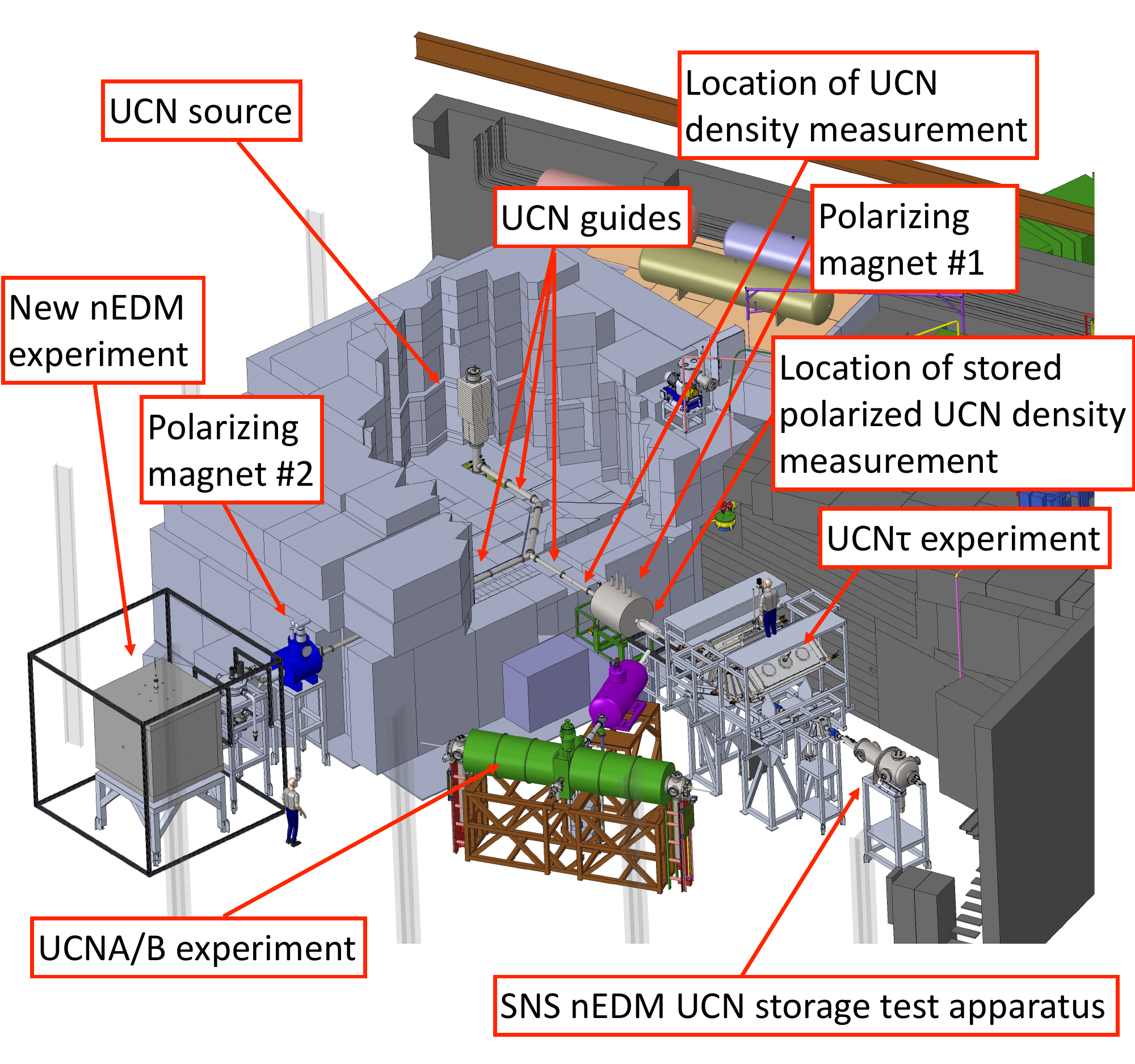}
\caption{The layout of the LANL UCN facility. Part of the biological
  shield is removed in this figure to show the UCN source and
  guides.\label{fig:AreaB}}
\end{figure}

\begin{figure}
\centering
\includegraphics[width=3.5in]{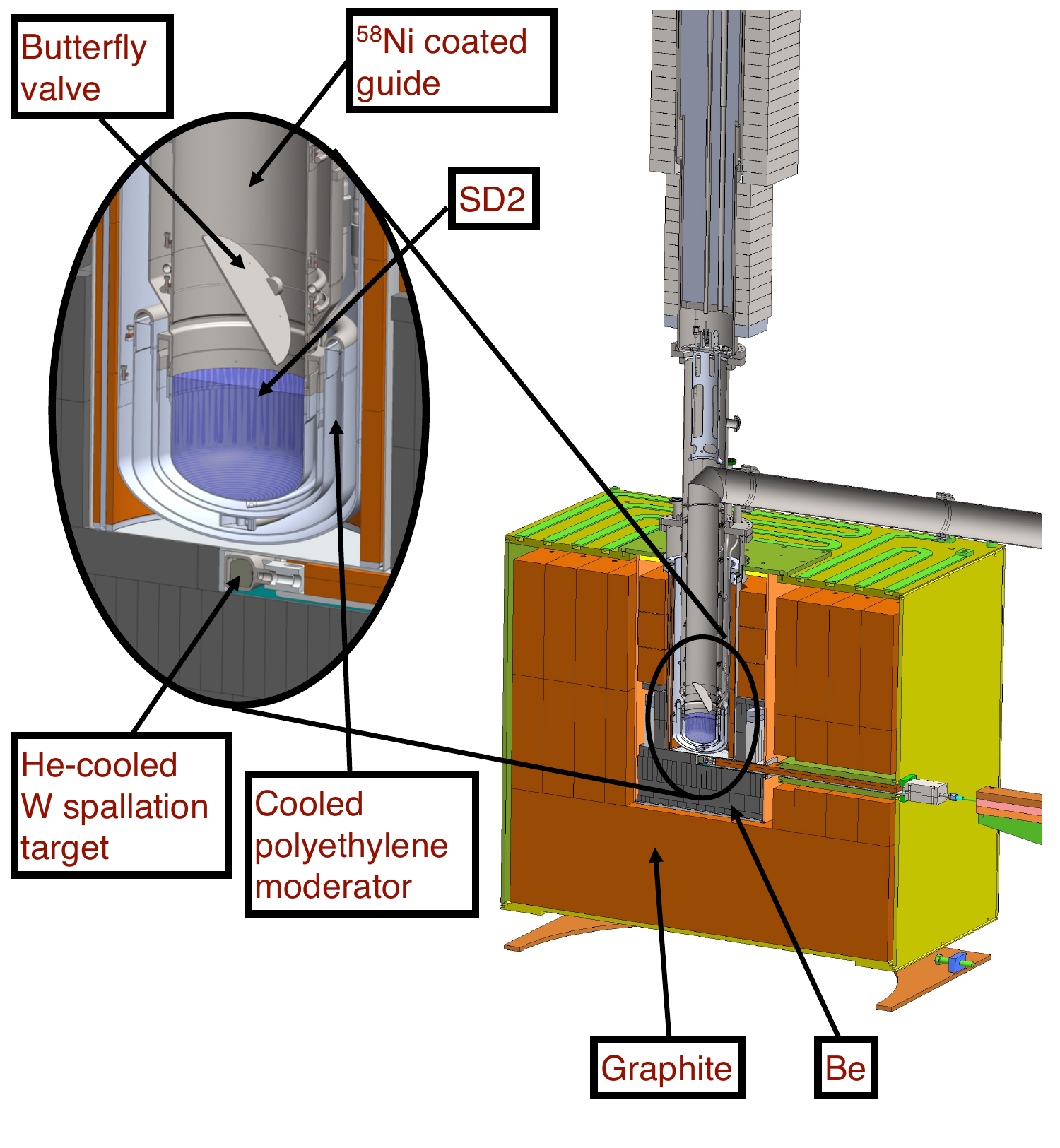}
\caption{Cutaway view of the source. The entire assembly is surrounded
  by the biological shield as shown in
  Fig.~\ref{fig:AreaB}.\label{fig:ucn_source}}
\end{figure}

Details of the design and performance of the LANL UCN source before
the upgrade are described in
Ref.~\cite{Sau13}. Figures~\ref{fig:AreaB} and \ref{fig:ucn_source}
show the layout of the LANL UCN facility and the details of the source
after the upgrade.

Because of the budgetary and schedule constraints, the scope of the
UCN source upgrade work was limited to replacing the so-called
``cryogenic insert'' and the horizontal UCN guide. The cryogenic
insert is the cryostat that houses the $^{58}$Ni-coated vertical
guide, the bottom of which is the SD$_2$ volume separated from the
rest of the vertical guide by the butterfly valve, and the cold
moderator volume. The scope of the upgrade work also included
installing an additional new UCN guide, which guides UCN to a location
envisioned for the new nEDM experiment (see Fig.~\ref{fig:AreaB}).

The design of the new cryogenic insert was optimized to maximize the
stored UCN density in the nEDM cell at the envisioned location of the
experiment. The optimization variables were the geometry of the
cryogenic insert, including the SD$_2$ volume, the cold moderator, and
the vertical and horizontal UCN guides, as well as the material and
temperature of the cold moderator. Considerations were given to both
the specific UCN production in the SD$_2$ volume and the UCN transport
from the SD$_2$ volume to the experiment. The specific UCN production
$P_{\rm UCN}$, the number of UCN produced per unit incident proton
beam charge per unit volume in the SD$_2$, is given by
\begin{equation}
P_{\rm UCN} = \rho_{\rm SD2} \int \Phi_{\rm CN}(E) \sigma_{\rm UCN}(E) dE,
\label{eq:ucnprod}
\end{equation}
where $\rho_{\rm SD2}$ is the number density of D$_2$ molecules,
$\Phi_{\rm CN}(E)$ is the cold neutron flux in the volume element
under consideration (per unit incident proton beam charge) at energy
$E$, and $\sigma_{\rm UCN}(E)$ is the UCN production cross section per
deuterium (D$_2$) molecule for cold neutrons at energy
$E$. $\sigma_{\rm UCN}(E)$ was taken from
Ref.~\cite{Fre10}. $\Phi_{\rm CN}(E)$ was evaluated using
MCNP6~\cite{MCNP6}. As seen in Fig.~6 of Ref.~\cite{Fre10},
$\sigma_{\rm UCN}(E)$ has a peak at $E\sim 6$~meV. The task was to
maximize the overlap integral in Eq.~(\ref{eq:ucnprod}). For the cold
moderator material, we considered polyethylene, liquid hydrogen, solid
and liquid methane, and mesitylene. The following thermal neutron
scattering cross section files were used in addition to the standard
MCNP6 distribution: Ref.~\cite{Gra09} for ortho-SD$_2$ at 5~K,
Ref.~\cite{Lav13} for polyethylene at 5, 77, and 293~K,
Ref.~\cite{Shi10} for solid methane at 20~K, and Ref.~\cite{Can06} for
mesitylene at 20~K. Polyethylene at 45~K, liquid methane at 100~K, and
mesitylene at 20~K gave equally good results. Solid methane at 20~K
made the cold neutron spectrum too cold. Taking engineering
consideration into account, we decided to use polyethylene beads
cooled to 45~K.

Our study showed that making the diameter of the SD$_2$ volume (and
the vertical guide and the opening of the cold moderator) smaller
increases the specific UCN production by confining the cold neutron
flux. However, the UCN transport study we performed using an in-house
developed UCN transport code indicated that by doing so, we lost in
UCN transport efficiency. As a compromise, we made the diameter of the
SD$_2$ volume somewhat smaller (15~cm) than that of the previous
source (20~cm). This allowed us to place the mechanism for the
butterfly valve outside the vertical guide, reducing the possible UCN
loss due to UCN interacting with potentially UCN-removing
surfaces. Furthermore, we increased the diameter of the horizontal
guide (up to the branch point at which the new UCN guide starts) to
15~cm from the previous 10~cm, further improving the UCN transport
efficiency. The horizontal UCN guides were all coated with nickel
phosphorus, which we measured to have a high Fermi potential [213(5)
  neV] and low UCN loss ($1.6\times 10^{-4}$ per
bounce)~\cite{Pat17b}.

The commissioning of the upgraded source started in November 2016. We
performed a series of measurements to characterize the performance of
the upgraded source. In order to make a direct comparison of the
performance of the upgraded source to that of the source before the
upgrade, these measurements were performed with the system
disconnected from the new UCN guide designed to send UCN to the new
nEDM experiment. The measured para fraction in the SD$_2$ was $2-3$\%
for these measurements.

The performance of SD$_2$-based UCN sources is known to depend on the
quality of the SD$_2$ crystal, as observed in our previous
source~\cite{Mak17} and others~\cite{Fre07,Lau16}. For the
measurements reported in this paper, the source was prepared following
our standard procedure, that is, the SD$_2$ crystal was grown directly
from D$_2$ gas at vapor pressures of $50-130$~mbar. With this source,
as well as with our previous source, we observed that the UCN
production degraded with time, and the crystal was therefore
periodically rebuilt. The source performance results presented in this
paper were typical.

We first measured the time of arrival (TOA) distribution of cold
neutrons at a detector placed 3.6~m above the 1-liter SD$_2$ volume, with the
timing of the proton pulse giving the start time. In order not to
blind the cold neutron detector (described below), these measurements
were performed with the proton beam pulsed at 1~Hz with each pulse
containing $\sim$1.5$\times 10^{-3}$~$\mu$C of protons.

\begin{figure}
\centering
\includegraphics[width=3.5in]{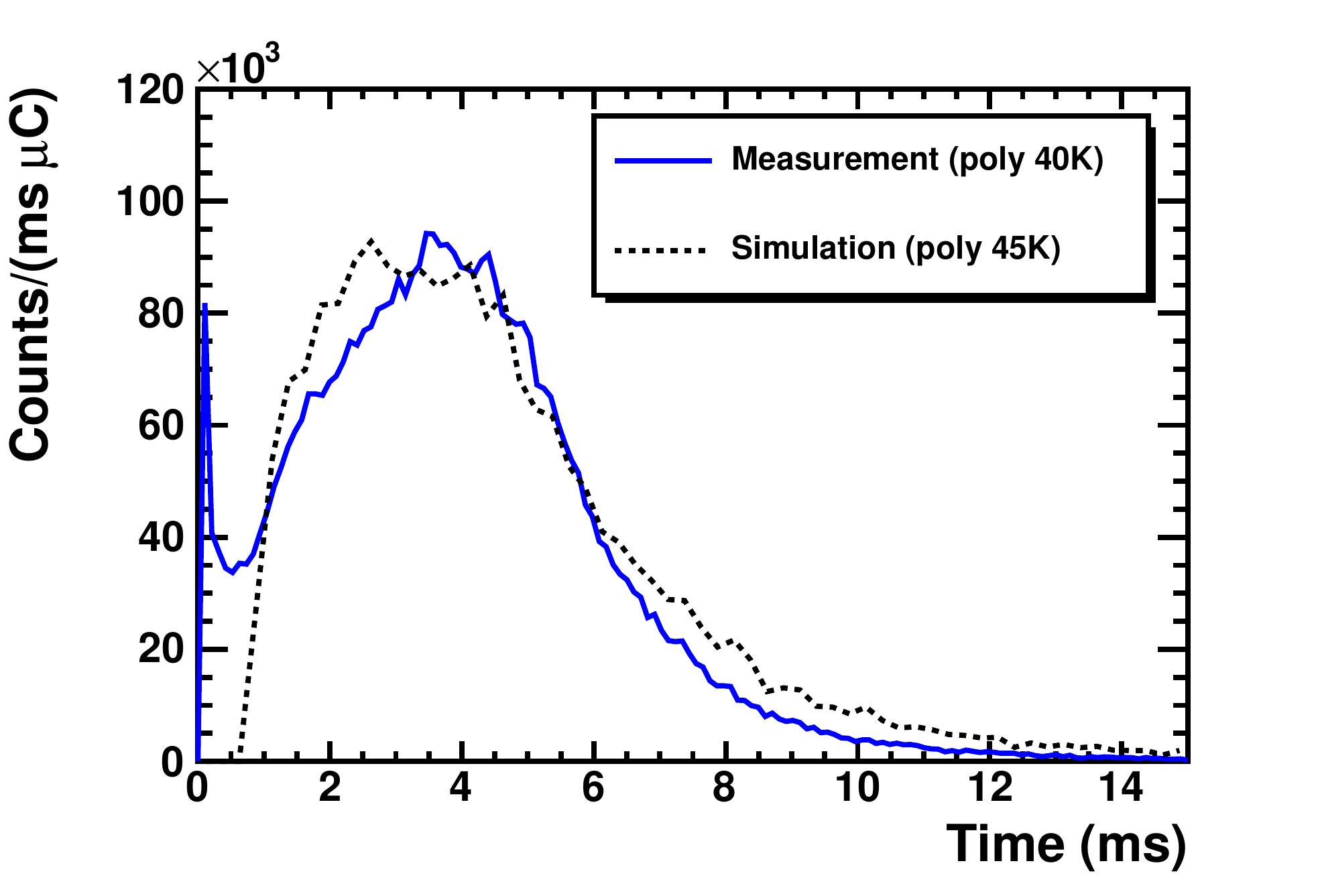}
\caption{Comparison between the measured cold neutron TOA spectrum and
  a simulated spectrum using MCNP6 for an SD$_2$ volume of
  1.1~liters. The simulation only includes neutrons with energies up
  to 100~meV. 
\label{fig:cn_tof}}
\end{figure}

The obtained TOA distributions give us information on the energy
spectra of the cold neutrons at the location of the SD$_2$, which we
can compare with our predictions from our MCNP6 model. If all cold
neutrons left the SD$_2$ volume at $t=0$, there would be a trivial
one-to-one correspondence between the neutron's energy and the TOA;
for example, it takes 3.4~ms for a 6-meV neutron to travel
3.6~m. However, our MCNP6 study showed that there is a broad emission
time distribution as is typical for cold neutron moderators. Therefore
a comparison between the experimental results and the simulations was
made using the TOA distributions, shown in Fig.~\ref{fig:cn_tof}. The
overall scale of the predicted spectrum was adjusted to reflect the
detection efficiency of the cold neutron detectors, which was measured
offline as described below. In the TOA range of $1.4-6.5$~ms, which
corresponds to neutron energies of $1.6-34$~meV in the absence of
delayed emission, the agreement between the measurement and model
prediction is better than 20\%, well within the expected model
uncertainty. However, the model clearly overpredicts for
TOA~$>6.5$~ms, the cause of which is a subject of further
investigation, along with a closer comparison between the measurement
and model for TOA~$<6.5$~ms. The same simulation predicts a specific
UCN production of 478~UCN/cm$^3$/$\mu$C near the bottom of the SD$_2$
volume and 290~UCN/cm$^3$/$\mu$C near the top of the SD$_2$ volume.

The cold neutron detector consisted of an Eljen-426HD2 scintillator
sheet directly coupled to a Hamamatsu R1355 photomultiplier tube
(PMT). The scintillator consisted of a homogeneous matrix of particles
of $^6$LiF and ZnS:Ag dispersed in a colorless binder~\cite{Eljen}. In
this scintillator sheet, while the $^6$LiF particles provided neutron
sensitivity through the neutron capture reaction
$^6$Li($n$,$\alpha$)$t$, the ZnS:Ag particles provided light output by
emitting scintillation light in response to alphas and tritons
traversing them. The alphas and tritons depositing part of their
energy in $^6$LiF particles or the binder produced a continuous energy
deposition spectrum, causing the detector efficiency to depend
strongly on the electronic threshold and to be lower than the
probability of neutron capture by $^6$Li. For the TOA measurement
describe above, the electronic threshold was adjusted so that the
detector was not blinded by the initial fast particles from the proton
pulse hitting the target. The efficiency of the cold neutron detector
with the same threshold setting was measured offline using a
monoenergetic neutron beam of 37~meV at the Neutron Powder Diffraction
Facility at the PULSTAR Reactor at North Carolina State
University. The detection efficiency was determined to be $0.53(2)$ of
the probability of neutron capture by $^6$Li in the sheet. The quoted
uncertainty is dominated by the systematic uncertainty associated with
the determination of the neutron flux.

We also measured the UCN density at two different locations using the
production proton beam described earlier. We first measured the
unpolarized UCN density at the exit of the biological shield (see
Fig.~\ref{fig:AreaB}). The measurement was performed using a small
vanadium foil fixed to the inner wall of the UCN guide, following the
method described in our earlier publication~\cite{Sau13}. The excited
$^{52}$V formed in the neutron capture reaction on $^{51}$V undergoes
$\beta$-decay to $^{52}$Cr and subsequently emits a 1.4-MeV $\gamma$
ray. At saturation, the rate of UCN capture equals the rate of 1.4-MeV
$\gamma$ emission, and the UCN capture rate can be related to the UCN
density by the kinetic theory formula $R=\frac{1}{4}\langle v \rangle
A\rho$, where $R$ is the capture rate, $A$ is the area of the foil,
$\langle v \rangle$ is the average UCN velocity, and $\rho$ is the UCN
density. A UCN transport Monte Carlo simulation study showed that the
UCN angular distribution is sufficiently isotropic at the location
where the density was measured and in the condition in which the
measurement was made to use this formula.  The $\gamma$ rays were
detected by a germanium detector placed outside the UCN guide. The
detection efficiency (including the detector solid angle) was
calibrated using a $^{60}$Co source of known activity placed at the
location of the vanadium foil. The UCN density was
$184(32)$~UCN/cm$^3$. The quoted uncertainty is dominated by the
systematic uncertainty associated with the correction for the effect
of the oxide layer, as described in Ref.~\cite{Sau13}.

\begin{figure}
\centering
\includegraphics[width=3.5in]{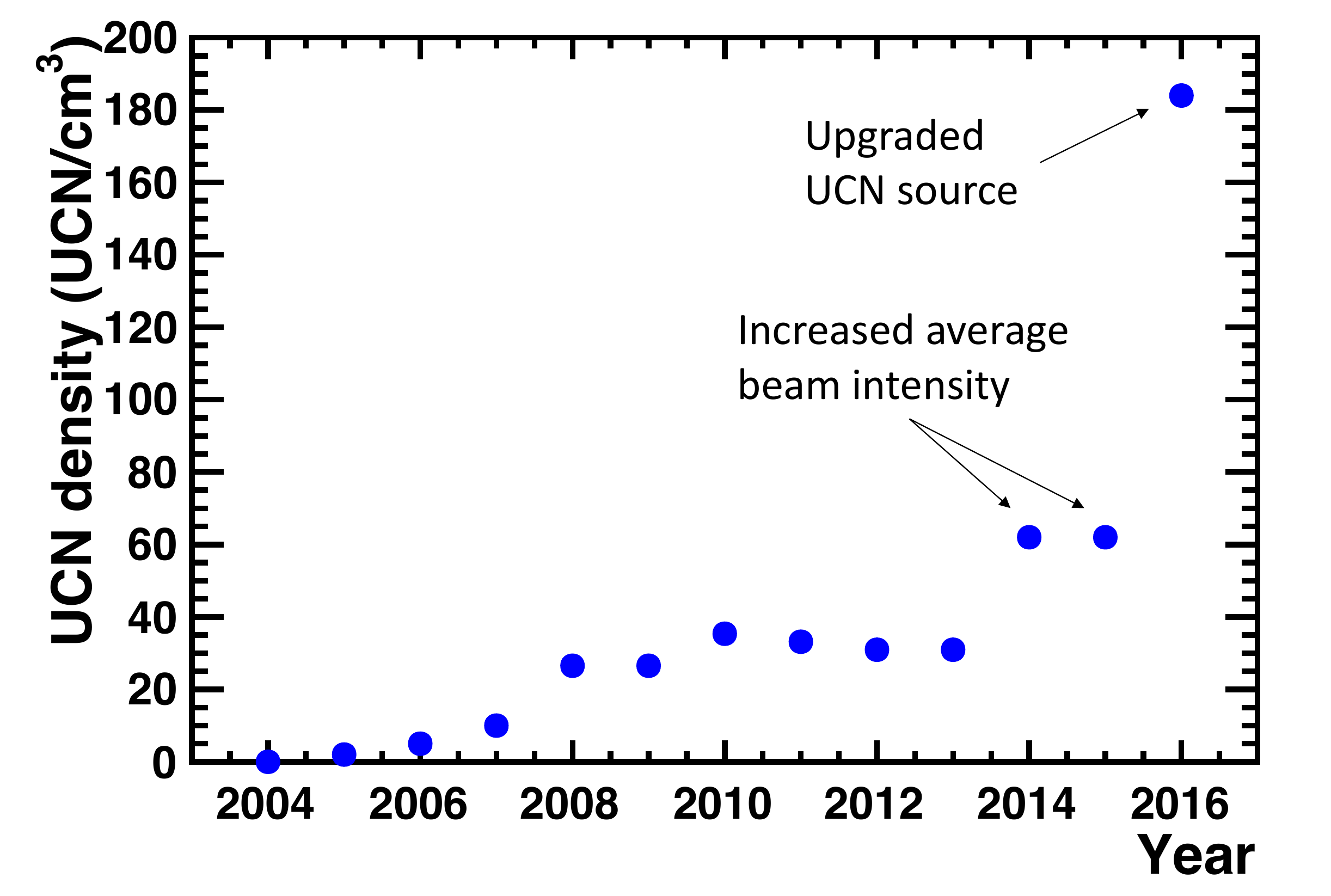}
\caption{UCN density from the LANL UCN source as measured at the exit
  of the biological shield as a function of year over the last 12
  years. The increase in 2014 was due to a new proton beam current
  monitor, which allowed us to run the proton beam closer to the
  allowed maximum of 10~$\mu$A. \label{fig:ucn_density}}
\end{figure}

Note that for this UCN density measurement, the UCN spectrum was cut
off by a stainless steel guide component used in the
system. Figure~\ref{fig:ucn_density} shows a comparison of this result
with ones from the previous source. Together with the higher average
proton current made possible by an improved proton beam current
monitor (a more accurate beam current integration allowed us to run
closer to the allowed limit), the UCN source upgrade project improved
the source performance relevant for filling the UCN chamber of a
neutron EDM experiment by a factor of $\sim$5.

\begin{figure}
\begin{center}
\includegraphics[width=3in]{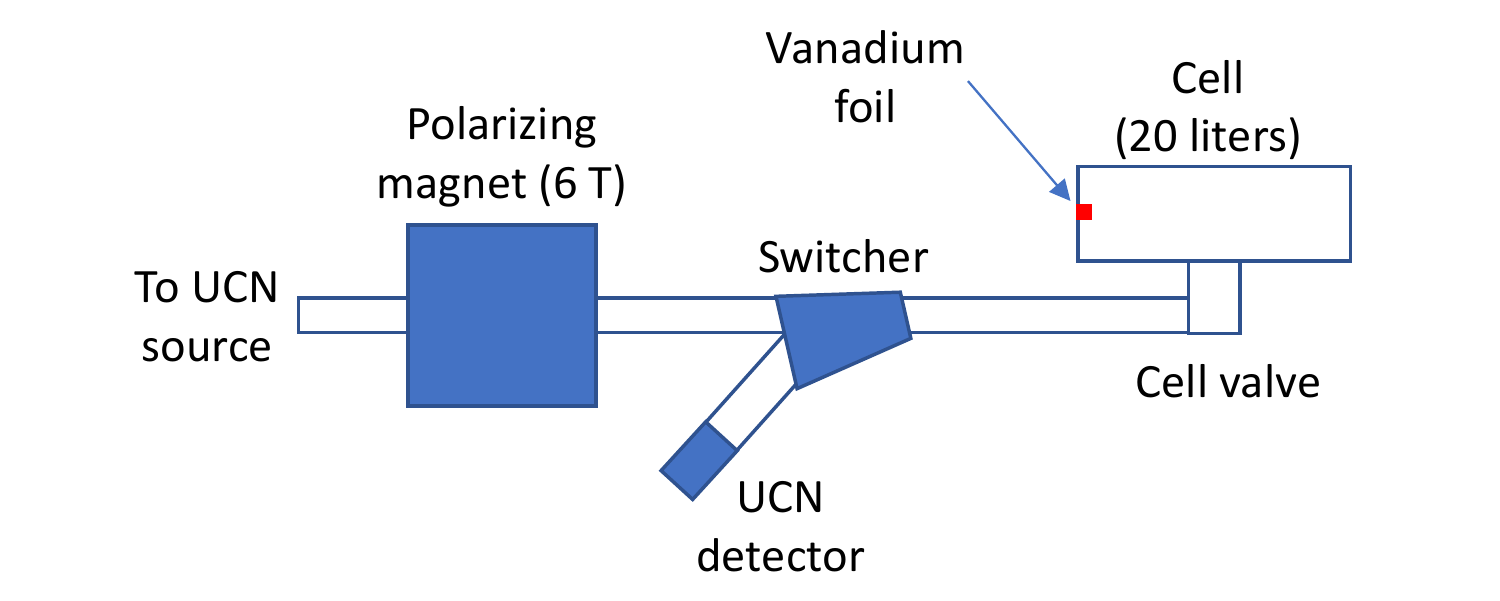}
\end{center}
\caption{A schematic of the experimental setup used
  for the measurement of the polarized UCN density stored in a
  prototype nEDM cell.\label{fig:schematic}}
\end{figure}

We measured the density of spin-polarized UCN stored in a prototype
nEDM cell located downstream of polarizing magnet \#1
(see~Fig.~\ref{fig:AreaB}). A schematic of the experimental setup is
shown in Fig.~\ref{fig:schematic}. The prototype nEDM cell had an
inner diameter of 50~cm and a height of 10~cm, giving a total volume
of 20~liters. The inner surface was coated with nickel
phosphorus. Note that for this UCN density measurement, in which high
field seeking UCN were stored in the cell, the UCN spectrum was cut
off by a stainless steel guide component used upstream in the system
and further softened by the $\sim$10~cm climb needed to enter the
chamber. The polarized UCN density was measured using the vanadium
foil method described above. The measured density at this location was
$39(7)$~UCN/cc. Here also, the quoted uncertainty is dominated by the
systematic uncertainty associated with the correction for the effect
of the oxide layer, as described in Ref.~\cite{Sau13}.

\begin{figure}
\begin{center}
\includegraphics[width=3.5in]{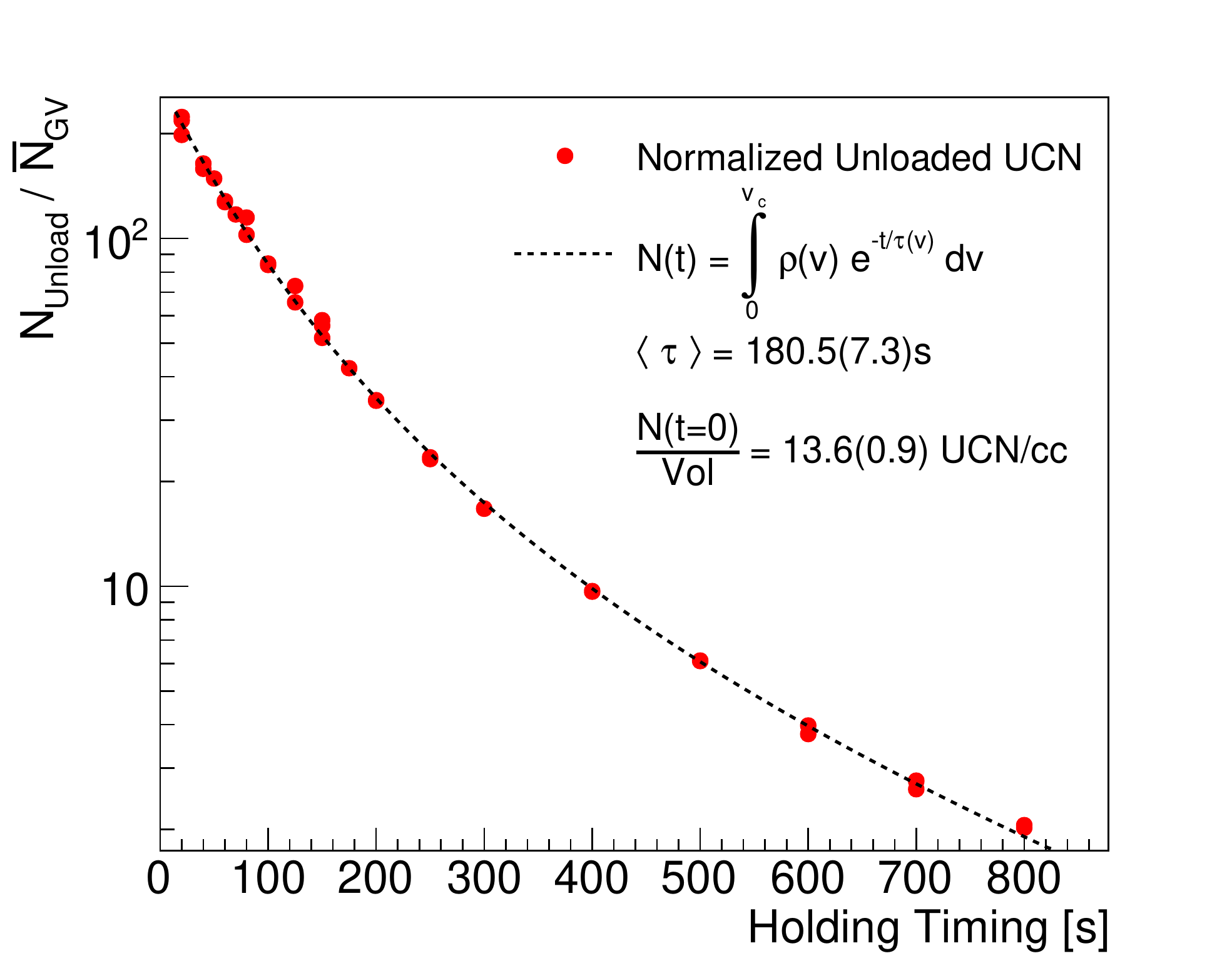}
\end{center}
\caption{The detected UCN counts normalized to the UCN monitor
  detector near the exit of the biological shield as a function of the
  holding time, in the fill-and-dump UCN storage measurement using our
  prototype nEDM cell.\label{fig:UCNstoragetime}}
\end{figure}

We also measured the stored spin-polarized UCN density in the
prototype nEDM cell using the so-called ``fill-and-dump'' method, in
which (i)~the cell is first filled with UCN, then (ii)~the cell valve
is closed to store UCN for a certain holding time, and finally
(iii)~the cell valve is opened and the remaining UCN are counted by a
detector mounted on a UCN switcher. This measurement is repeated for
various holding times. After holding times of 20~s and 150~s, we
detected $\sim$200,000 UCN and $\sim$45,000 UCN,
respectively. Figure~\ref{fig:UCNstoragetime} shows the detected UCN
counts, normalized to the UCN monitor detector near the exit of the
biological shield, as a function of the holding time. Normalization
was necessary to make a storage time curve because the UCN source
output had some fluctuations because the proton bursts from the
accelerator were separated by 10~s. The stored UCN density
extrapolated to $t=0$ was $13.6(0.9)$ UCN/cc, with the quoted
uncertainty dominated by the fluctuation of the UCN source output
mentioned above. The difference between this density and the density
obtained by the vanadium method can be attributed to the loss of UCN
through the UCN switcher as well as the finite UCN detection
efficiency. The curve in Fig.~\ref{fig:UCNstoragetime} is well
described by the UCN spectrum having the form $\frac{dN}{dv}\propto
v^{2.9(0.6)}$ with a cutoff velocity of 5.7~m/s and the cell having an
average storage time of 181(7)~s.



The system lifetime of the UCN source with the butterfly valve closed was
measured to be 84.2(1.9)~s. 

Using the measured spin polarized UCN density stored in the prototype
nEDM cell, we can estimate the statistical sensitivity of a possible
nEDM experiment mounted at the upgraded UCN source. When an nEDM
experiment based on Ramsey's separated oscillatory field
method~\cite{Ram56} is performed using stored UCN, the statistical
sensitivity is given by
\begin{equation}
\delta d_n = \frac{\hbar}{2\alpha E T_{\rm free}
  \sqrt{N_T}}, 
\end{equation}
where $\alpha$ is the polarization product (the product of the
analyzing power and the UCN polarization), $E$ is the strength of the
electric field, $T_{\rm free}$ is the free precession time, and $N_T$
is the total number of the detected neutrons over the duration of the
experiment.  If the number of detected neutron per measurement cycle
is given by $N$, and the number of the performed measurement cycles is
given by $M$, then $N_T = NM$. Therefore, the cycle time $T_{\rm
  cycle}$, the time it takes to perform a measurement, is also an
important parameter. For the purpose here, we use the detected number
of UCN at 180~s holding time, which is 39,000 per cell, for $N$. We
further assume $\alpha=0.8$, $E=12$~kV/cm, $T_{\rm free}=180$~s, and
$T_{\rm cycle}=300$~s, all of which have been demonstrated by other
collaborations (see e.g. Ref.~\cite{Bon16}). Furthermore, we assume a
double chamber configuration~\cite{Ser15}. Then we expect to achieve a
per-day one-standard-deviation statistical sensitivity of $\delta d_n
= 4.0\times 10^{-26}$~$e\cdot$cm/day. With an assumed data taking
efficiency of 50\% and the nominal LANSCE accelerator running
schedule, we expect to achieve a one standard deviation statistical
sensitivity of $\delta d_n = 2.1\times 10^{-27}$~$e\cdot$cm in 5
calendar years. If the systematic uncertainty is equal to the
statistical, then the total one-standard-deviation sensitivity is
$\sigma(d_n) = 3\times 10^{-27}$~$e\cdot$cm. Note that this is a
conservative estimate, as it is expected that we can further improve
the detected number of UCN by using a switcher with an improved
design, which will result in fewer years needed to achieve this
sensitivity.

In conclusion, we have successfully upgraded the LANL UCN source. Its
performance, as evaluated by the measured equilibrium UCN density in
the UCN guide at the exit of the biological shield [184(32)~/cm$^3$]
and the CN TOA distributions measured by a CN detector located 3.6~m
above the source, is consistent with the prediction based on MCNP6
neutron moderation simulation and an in-house UCN transport
simulation. The measured polarized UCN density stored in an nEDM-like
chamber indicate that an nEDM experiment based on the room temperature
Ramsey's separated oscillatory field method with a
one-standard-deviation sensitivity of $\sigma(d_n)=3\times
10^{-27}$~$e\cdot$cm is possible with a running time of five calendar
years (possibly fewer years with the switcher transmission issues
addressed). In addition, with the upgraded LANL UCN source, the
UCN$\tau$ experiment now routinely collects data sufficient for a 1-s
statistical uncertainty in the free neutron lifetime in an actual
running time of $\sim$60 hours~\cite{Pat17}. The upgraded LANL UCN
source will enable other UCN based experiments such as improved
measurements of the neutron $\beta$-asymmetry.

This work was supported by Los Alamos National Laboratory LDRD Program
(Project No. 20140015DR), the US Department of Energy (contract
numbers DE-AC52-06NA25396 and DE-FG02-97ER41042) and the US National
Science Foundation (Grant No. PHY1615153). We gratefully acknowledge
the support provided by the LANL Physics and AOT Divisions. We thank
M.~Mocko and his team for allowing us to use their computer cluster
for the mcnp6 simulation presented in this paper. We thank E.~Brosha,
S.~Satjija, and G.~Yuan for their help with characterizing the
$^{58}$Ni coating.


\begin{thebibliography}{99}
\bibitem{Ign90}
V.~K.~Ignatovich, {\it The Physics of Ultracold Neutrons} (Clarendon,
Oxford, 1990).

\bibitem{Gol91}
R.~Golub, D.~Richardson, S.~K.~Lamoreaux, {\it Ultra-Cold Neutrons}
(Adam Hilger, Bristol, 1991). 

\bibitem{Dub11}
D.~Dubbers and M.~G.~Schmidt, Rev. Mod. Phys. {\bf 83,} 1111 (2011).

\bibitem{You14}
A.~R.~Young, {\it et al.} J. Phys. G: Nucl. Part. Phys. {\bf 41,}
114007 (2014).

\bibitem{Khr97}
I.~B.~Khriplovich and S.~K.~Lamoreaux, {\it CP Violation Without
  Strangeness} (Springer-Verlag, Berlin 1997).

\bibitem{Lam09}
S.~K.~Lamoreaux and R.~Golub, J. Phys. G: Nucl. Part. Phys. {\bf 36,}
104002 (2009).

\bibitem{ILL}
C.~A.~Baker, {\it et al.}, Phys. Rev. Lett. {\bf 97,} 131801 (2006);
J.~M.~Pendlebury, {\it et al.}, Phys. Rev. D {\bf 92,} 092003 (2015).

\bibitem{Ser15}
A.~P.~Serebrov, {\it et al.}, Phys. Rev. C {\bf 92}, 055501 (2015).

\bibitem{Pos05}
M.~Pospelov and A.~Ritz, Ann. Phys. (NY) {\bf 318}, 119 (2005).

\bibitem{Eng13}
J.~Engel, M.~J.~Ramsey-Musolf, and U. van Kolck,
Prog. Part. Nucl. Phys. {\bf 71}, 21 (2013).

\bibitem{Cir10} 
V.~Cirigliano, Y.~Li, S.~Profumo, and M.~J.~Ramsey-Musolf, JHEP {\bf
  2010,} 2 (2010).

\bibitem{Mor13}
D.~E.~Morrissey and M.~J.~Ramsey-Musolf, New J. Phys. {\bf 14}, 125003 (2012).

\bibitem{Ser08}
A.~Serebrov {\it et al.}, Phys. Rev. C {\bf 78,} 035505 (2008).

\bibitem{Pat17}
R.~W.~Pattie Jr. {\it et al.} (UCN$\tau$ Collaboration), {\tt arxiv:1701.01817}.

\bibitem{Yue13}
A.~Yue {\it et al.}, Phys. Rev. Lett. {\bf 111,} 222501 (2013).

\bibitem{Pat09}
R.~W.~Pattie, Jr. {\it et al.}, Phys. Rev. Lett. {\bf 102,} 012301 (2009).

\bibitem{Liu10}
J.~Liu {\it et al.}, Phys. Rev. Lett. {\bf 105,} 181803 (2010).

\bibitem{Pla12}
B.~Plaster {\it et al.} (UCNA Collaboration), Phys. Rev. C {\bf 86,}
055501 (2012).

\bibitem{Men13} 
M.~P.~Mendenhall {\it et al.} (UCNA Collaboration), Phys. Rev. C {\bf
  87,} 032501(R) (2013).

\bibitem{Bro17}
  M.~A.-P.~Brown {\it et al.} (UCNA Collaboration), Phys. Rev. C
  (accepted for publication) [arXiv:1712.00884].

\bibitem{Mun13}
D.~Mund {\it et al.}, Phys. Rev. Lett. {110}, 172502 (2013).

\bibitem{Cir13}
V.~Cirigliano, S.~Gardner, B.~R.~Holstein,
Prog. Part. Nucl. Phys. {\bf 71,} 93 (2013).

\bibitem{Jen11} 
T.~Jenke, P.~Geltenbort, H.~Lemmel, H.~Abele, Nat. Phys. {\bf 7,} 468
(2011).

\bibitem{Ste86}
A.~Steyerl {\it et al.} Phys. Lett. A {\bf 116,} 347 (1986).

\bibitem{Gol75}
R.~Golub and J.~M.~Pendlebury, Phys. Lett. {\bf 53A,} 133 (1975).

\bibitem{Gol77}
R.~Golub and J.~M.~Pendlebury, Phys. Lett. {\bf 62A,} 337 (1977).

\bibitem{Gol83}
R.~Golub and K.~B\"oning, Z. Phys. B {\bf 51,} 95 (1983).

\bibitem{Yu85}
Z.-Ch.~Yu, S.~S.~Malik, and R.~Golub, Z. Phys. B {\bf 62,} 137 (1985).

\bibitem{Mor03}
C.~L.~Morris {\it et al.}, Phys. Rev. Lett. {\bf 89,} 272501 (2002).

\bibitem{Sau04}
A.~Saunders {\it et al.}, Phys. Lett. B {\bf 593,} 44 (2004).

\bibitem{Sau13}
A.~Saunders {\it et al.}, Rev. Sci. Instrum. {\bf 84,} 013304 (2013).

\bibitem{Bro16}
L.~J.~Broussard {\it et al.}, Nucl. Instrum. Methods
Phys. Res. Sect. A {\bf 849,} 83 (2017).

\bibitem{Tan16}
Z.~Tang at {\it International Workshop: Probing Fundamental
  Symmetries with UCN}, April 2016, Mainz, Germany (2016).

\bibitem{Ito14}
Los Alamos National Laboratory LDRD Project \#20140015DR, ``Probing New
Sources of Time-Reversal Violation with Neutron EDM'', T.~M.~Ito, PI.

\bibitem{Fre10}
A.~Frei, E.~Gutsmiedl, C.~Morkel, A.~R.~M\"uller, S.~Paul, S.~Rols,
H.~Schober, and T.~Unruh, Europhys. Lett. {\bf 92,} 62001 (2010).

\bibitem{MCNP6}
Los Alamos National Laboratory Monte Carlo Code Group, `` A General
Monte Carlo N-Particle (MCNP) Transport Code'', {\tt
  http://mcnp.lanl.gov}

\bibitem{Gra09}
J.~R.~Granada, Eur. Phys. Lett. {\bf 86,} 66007 (2009).

\bibitem{Lav13}
C.~M.~Lavelle, C.~Y.~Liu, M.~B.~Stone,
Nucl. Instrum. Methods Phys. Res. A {\bf 711,} 166 (2013).

\bibitem{Shi10}
Y.~Shin, W.~M.~Snow, C.~Y.~Liu, C.~M.~Lavelle, D.~V.~Baxter,
Nucl. Instrum. Methods Phys. Res. A {\bf 620,} 382 (2010).

\bibitem{Can06} F.~Cantargi, J.~R.~Granada, S.~Petriw, M.~M.~Sbaffoni,
  Physica B {\bf 385,} 1312 (2006).

\bibitem{Liu00}
C.-Y.~Liu, A.~R.~Young, and S.~K.~Lamoreaux, Phys. Rev. B {\bf 62,}
R3581 (2000)

\bibitem{Pat17b}
R.~W.~Pattie Jr. {\it et al.}, Nucl. Instrum. and Methods
Phys. Res. Sect. A {\bf 872,} 64 (2017).

\bibitem{Mak17}
M~.Makela (private communications).

\bibitem{Fre07}
A.~Frei {\it et al.}, Eur. Phys. J. A {\bf 34}, 119 (2007).

\bibitem{Lau16}
B.~Lauss at {\it International Workshop: Probing Fundamental
  Symmetries with UCN}, April 2016, Mainz, Germany (2016).

\bibitem{Eljen}
Eljen Technology, {\tt http://www.eljentechnology.com}.

\bibitem{Ram56}
N.~F.~Ramsey, {\it Molecular Beams} (Oxford University Press, Oxford,
1956).

\bibitem{Bon16}
V.~Bonder at {\it International Workshop: Probing Fundamental
  Symmetries with UCN}, April 2016, Mainz, Germany (2016).


\end{thebibliography}
\end{document}